\documentclass[floatfix,showpacs,aps,prd,amsmath,amssymb,twocolumn,10pt,sort&compress]{revtex4}
\usepackage{amsmath}
\usepackage{epsfig}
\usepackage{subfigure}

\newcommand{\be}{\begin{equation}}\newcommand{\ee}{\end{equation}}
\newcommand{\bea}{\begin{eqnarray}}\newcommand{\eea}{\end{eqnarray}}
\newcommand{\bc}{\begin{center}}\newcommand{\ec}{\end{center}}

\def\lsim{\raise0.3ex\hbox{$\;<$\kern-0.75em\raise-1.1ex\hbox{$\sim\;$}}}
\def\gsim{\raise0.3ex\hbox{$\;>$\kern-0.75em\raise-1.1ex\hbox{$\sim\;$}}}

\begin{document}

\bigskip
\title{ A Simple Expression for Heavy to Light Meson 
\\Semileptonic  Decays Form Factors}

\author{T. N. Pham\footnote{email address: pham@cpht.polytechnique.fr}}

\affiliation{Centre de Physique Th\'{e}orique, CNRS,Universit\' e  Paris-Saclay \\ 
Ecole Polytechnique, 91128 Palaiseau, Cedex, France }
\date{\today}

\begin{abstract}
\parbox{14cm}{\rm 
  
  Like the two-photon and  two-gluon decays of the
$P$-wave $\chi_{c0,2}$ and $\chi_{b0,2}$ charmonium state for which  
the  Born term produces a very simple decgays amplitude in terms of 
an effective Lagrangian with two-quark local operator, the   Born term
for  the processes  $c\bar{d} \to  (\pi,K) \ell \nu $ and $b \bar{d}\to (\pi,
K) \ell\nu $, could also  produce the  $D$ and $B$ meson semileptonic 
decays with the light  meson $\pi, K$ in the final state. In  this  
approach to heavy-light meson form factors with the $\pi$, K meson treated as
Goldstone boson, a simple  expression is found for the decays 
form factors, given  as~:  $ f_{+}(0)/(1 -q^{2}/(m_{H}^{2}+ m_{\pi}^{2}))$, 
 with $H=D,B$ for $D,B \to \pi$  form factors, and 
  $f_{+}(0)/(1 -q^{2}/(m_{H}^{2}+ m_{K}^{2}))$ for  $B,D\to K$ form factor.
The purpose of this paper is to show that this expression  for the form 
factors could  describe rather well the $q^{2}$- behaviour observed in 
 the BaBar,  Belle and BESIII  measurements and lattice simulation. In 
particular,  the  $D\to K$  form factors are in good agreement with the
 measured values in the whole range of $q^{2}$ showing evidence for
 $SU(3)$ breaking with the presence of $m_{K}^{2}$ term in the quark 
propagator, but some  corrections  to the Born term are needed at 
large $q^{2}$ for $D,B \to \pi$  form factors.}

\end{abstract}
\pacs{13.25.Hw 11.10.Hi 13.28.Bx}
\maketitle

\section{Introduction}
The  semileptonic heavy to light meson decays form factor as in  the
$D,B \to \pi \ell \nu$ decays, is given by the   
$V-A$ current matrix elements between heavy and light meson state. The precise
knowledge of these form factors is required  for an exclusive
determination  of the CKM matrix element $V_{ub}$. There have been
previous calculations of the form factors in the light-cone sum rule
approach(LCSR)\cite{Ball} , in  lattice simulation\cite{Bailey,Al-Haydari},
and are  known experimentally from  BaBar\cite{BaBar}, Belle\cite{Belle}
and BESIII\cite{BES}  measurements. It  is shown that the 
 $B\to \pi$ form factor from lattice simulation and BaBar measurements could be
fitted with a BK parametrization\cite{BK}  with 
$\alpha=0.52 \pm 0.05\pm 0.03$ given in \cite{BaBar1}
\be
f^{+}(q^{2},\alpha) = \frac{f_{0}}{(1 -q^{2}/m^{2}_{B^{*}})(1 -\alpha q^{2}/m^{2}_{B^{*}})}
\label{fpB}
\ee 
  For $D$ meson semileptonic   decays, the BaBar and BESIII 
measurements\cite{BES,BaBar2}  show that the  $D\to \pi, K$ form factor , could be fitted with an effective  two-pole~:
\be
f^{+}(q^{2},\alpha) = \frac{f_{0}}{(1 -q^{2}/m^{2}_{D^{*}}).(1 -\alpha q^{2}/m^{2}_{D^{*}})}
\label{fpD}
\ee
or a single-pole parametrization~:
\be
f^{+}(q^{2},\alpha) = \frac{f_{0}}{(1 -q^{2}/m^{2}_{\rm pole})}
\label{fpS}
\ee
with  $m^{2}_{\rm pole}=(1.906 \pm 0.029\pm 0.023){\rm GeV}$
in the BaBar measurements\cite{BaBar2} for $D\to \pi$ form factor and in  
the BESIII measurements, 
 $m_{\rm pole}= (1.911 \pm 0.012\pm 0.004){\rm GeV}$
for $D\to \pi$, and $m_{\rm pole}= (1.921 \pm 0.010\pm 0.007){\rm
  GeV/}$ for $D\to K$ form factor which seem to disagree with the 
higher pole mass value given by $D^{*}$ dominance as noted by BESIII 
\cite{BES}. The lower value for the pole mass is also found by 
Belle\cite{Belle} which give a value of  $1.82\pm 0.04\pm 0.03 \rm GeV$,
but a higher value for pole mass of  $(1.93 \pm 0.05\pm 0.03){\rm  GeV}$
close to the BESIII value is obtained  earlier\cite{FOCUS}. Other data
for  $D\to K$ form factor are given in TABLE VI of Ref. [\cite{BaBar3}],
with the pole mass values for CLEO and FOCUS around $1.9 {\rm  GeV}$ close
to the BESIII values.

  The problem is to obtain a theoretical expression with this pole-dominance
 $q^{2}$-behaviour for these form factors. As will be shown in the
following, it is quite straightforward to obtain the pole 
dominance term for the form factors by noting that the form factor for
semileptonic decay with pion or light hadron in the final state is, at the
quark level, a $c\bar{d} \to\pi \ell \nu $ and $b \bar{d}\to \pi \ell\nu $ 
process, with the light pseudo scalar meson $\pi, K$  treated as the 
Goldstone boson of the chiral $SU(3)\times SU(3)$ symmetry, rather
than a two-body bound state which gives a rather small 
$\bar{B^{0}}\to \pi^{+}\ell^{-}\nu$ branching ratio\cite{Suzuki1}. Like  the 
two-photon and  two-gluon decays of the $P$-wave  $\chi_{c0,2}$ and 
$\chi_{b0,2}$ charmonium  state\cite{Barbieri,Kuhn,Lansberg}, the  Born
term from  this process then gives the semileptonic decay amplitude 
and the  $D \to \pi$ and $B\to \pi$  form factors in terms of  the light 
pseudo  scalar meson-quark  coupling $g_{\pi qq}$  given by  the
 Goldberger-Treiman relation\cite{Pham}, the  $c\bar{d} \to\pi \ell \nu $ and $b \bar{d}\to \pi \ell\nu$ process  then produce the form factors.
The use of the pion-quark coupling to obtain the semileptonic decay amplitudes
is similar    to  the study of strong and radiative
decays of vector mesons with an essentially the same  Born term for the 
radiative decay  $\rho^{-} \to \pi^{-}\gamma$ from which the extracted 
pion-quark coupling  consistent  with the theoretical value given by the bag 
model\cite{Suzuki2} to within $50\%$ . Thus it is possible to obtain the
semileptonic decays amplitude with pion in the final state treated as a 
Goldstone  boson of chiral symmetry with the pion-quark coupling given 
by the Goldberger-Treiman relation.  The purpose of this
paper is to show that, from a  similar Born term , we could 
obtain  the   semileptonic decay form factors of $B,D$ meson with a
light pseudo-scalar meson in the  final states from the Born term for the 
 $c\bar{d} \to\pi \ell \nu $ and $b \bar{d}\to \pi \ell\nu$ process. We
 find  that the  $q^{2}$-dependence given by the Born term
describes quite well the form factor  measurements by BaBar, Belle and
BESIII experiments. In  particular, the value of the BESIII  fitted pole  
mass\cite{BES,BES2} in $D\to K \ell\nu$ is very close to the effective pole
mass of the Born term, showing  the  presence of the $m_{K}^{2}$ term in 
the quark propagator.

\section{Effective Lagrangian for $D\to \pi \ell\nu$ and $B\to \pi \ell\nu$}

 Like the $c\bar{c}\to \gamma\gamma$  annihilation in the two-photon decay
of $P-$wave charmonium state which, at the tree-level approximation, 
proceeds through the Born term, the reactions 
  $c+ {\bar d}\to \pi  \ell \nu$ and $b+ {\bar d}\to \pi  \ell \nu$, can
occur through a similar Born term, with the exchange of an $u$ quark 
which combines with $\bar{d}$ quark to produce a pion in the final 
state. We have:
\be
 O_{\mu} = \bar{v}(p_{d})V_{\mu} u(p_{b})
\label{amp}
\ee
with 
\bea
 V_{\mu} =\frac{1}{i}\Biggl[(-ig_{\pi qq}\gamma_{5})
i\frac{(\not\! p-\not\! p_{d} +m_{u})}{(p-p_{d})^{2}
  -m_{u}^{2}}(-ig\gamma_{\mu})\Biggr] 
\label{Vpi}
\eea
with $p_{b}$ and $p_{d}$  the  $b$ and $\bar{d}$ momenta of the $B$ 
meson with mass $m_{B}$ and $q$ and $p$ are respectively the final state electron-antineutrino
system and the pion momentum as shown in FIG.1

\begin{figure}[ht]
\centering
\includegraphics[height=3.0cm,angle=0]{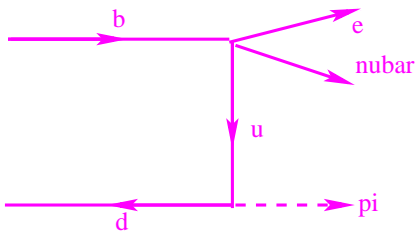}
\caption{ Diagram showing  Born term for the semileptonic  B decay form factor}
\label{fig1}
\end{figure}.
   For the quark mass in the quark propagator, we use the constituent 
quark mass given in\cite{Ali}, with
\bea
&&  m_{b}= 4.88 \ {\rm GeV}, m_{c}= 1.5 \ {\rm  GeV} \nonumber\\
&&  m_{s}= 0.5 \ {\rm GeV}, m_{u}= m_{d}= 0.3 \ { \rm  GeV}
\label{mass}
\eea
   To obtain the form factor, as with 
 $B_{s}\to \gamma \gamma$ decay\cite{Ricciardi}, we  work  in the  
weak binding approximation with  $b,\bar{d}$ quark taken at rest in 
the  $B^{0}$ meson. This immediately gives us the effective
Lagrangian for $B, D \to \pi \ell \nu$. Using the Dirac equation and
letting the  $b$ and $\bar{d}$ quark on the mass shell, with
    $m_{d}=m_{u}$, we have :
\bea
 V_{\mu} =\frac{1}{i}\Biggl[(-ig_{\pi qq}\gamma_{5})
i\frac{\not\! p}{(p^{2} -2m_{q}p_{0})
  }(-ig\gamma_{\mu})\Biggr] 
\label{Vpi2}
\eea
Putting $V_{\mu}$ into $(O_{1})$ in Eq. (\ref{amp}) and replacing 
$\bar{v}(p_{d})$ and $u(p_{b})$ with the quark field $\bar{u}$ and
$b$  in $(O_{1})$, we obtain the following local operator for the 
vector current matrix element in $B\to \pi \ell \nu$ decays:
\be
 O_{P} = \frac{2m_{B}(\bar{d}\gamma_{5} b)p_{\mu} }{(m_{B}^{2}
 +m_{\pi}^{2} - q^{2})}
\label{Ops} 
\ee
showing the appearance of the pole at $q^{2}= (m_{B}^{2} +m_{\pi}^{2}) $ 
generated by the Born term.   For $D\to K$ form factor, the pole  is 
at $q^{2}= (m_{D}^{2} +m_{K}^{2}) $. This result explains the
success of the single-pole or two-pole fits of the BaBar and BESSIII data
as shown below. With
\be
 <\pi(p)|V_{\mu}(0)|B(p_{B})> = f_{+}(q^{2})(p_{B} + p)_{\mu}
\label{ffp}
\ee
and with $m_{b}+ m_{d}= m_{B}$, $<0|\bar{d}\gamma_{5}b|B)= m_{B}f_{B}$,
the matrix element $<0|O_{P}|B>$ then gives us immediately the form factor 
for  $B\to \pi \ell \nu$ decay. 

  Thus by putting the heavy quark at rest, we are able to obtain the effective
Lagrangian for  the semileptonic decays $B, D \to \pi \ell \nu$ which
then gives us the form factors in a very simple expression. This is not
the case for the radiative decays of light vector meson, e.g the 
$\rho$ meson,  for which the exact calculation given in \cite{Leinweber}
is also obtained with expressions far  more involved than the simple
result in  the static approximation mentioned earlier\cite{Suzuki2}.

  Using the  pion-quark coupling for a constituent quark in the $B$ meson
obtained from the Goldberger-Treiman relation for the pion-nucleon 
coupling constant with $g_{A}=3/4 $ \cite{Pham} and dropping the weak
coupling  constant $g$ not relevant for our purpose, we have:
\be
\kern -0.5cm   f_{+}(q^{2})= \biggl(\frac{f_{B}}{f_{\pi}}\biggr)g_{A}
\frac{1}{(1 + m_{\pi}^{2}/m_{B}^{2})}\frac{1}{(1 -q^{2}/(m_{B}^{2}+m_{\pi}^{2}))}
\label{ffB}
\ee
for $q^{2}=0$, we have:
\be
\kern -0.5cm   f_{+}(0)= \biggl(\frac{f_{B}}{f_{\pi}}\biggr) g_{A}\frac{1}{(1 + m_{\pi}^{2}/m_{B}^{2})}
\label{ffB0}
\ee
 and similar expression for $D\to K$ form factor.

\begin{figure}[ht]
\centering
\includegraphics[height=5cm,angle=0]{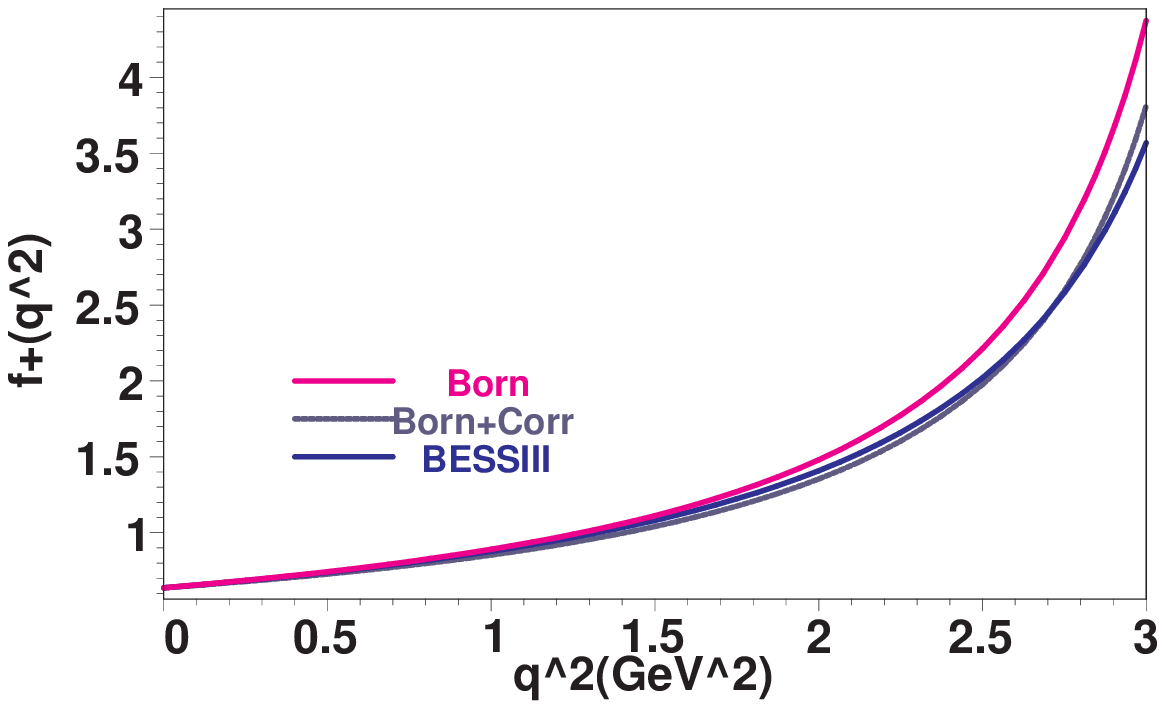}
\caption{The Born term(upper curve), the Born term with a small
polynomial term to fit the $D^{0}\to \pi^{-}$ form factor BESIII data
(lower curve) which are   given in FIG.9 of Ref. \cite{BES}}.
\label{fig2}
\end{figure}

\begin{figure}[ht]
\centering
\includegraphics[height=5cm,angle=0]{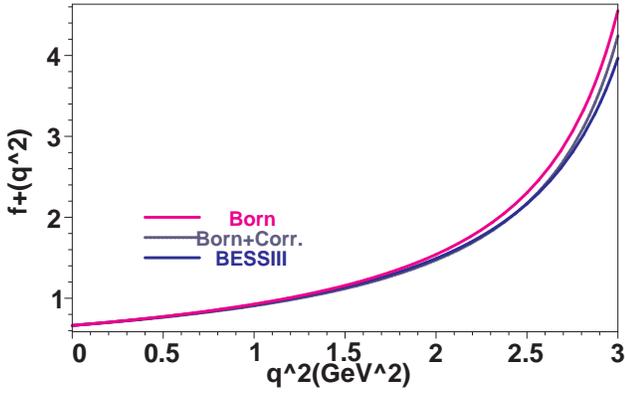}
\caption{The Born term(upper curve), the Born term with a small
polynomial term to fit the $D^{+}\to \pi^{0}$ form factor BESIII data
lower curve) which are  given in FIG.6 of Ref. \cite{BES2}}
\label{fig3}
\end{figure}

There  are also possible suppression of the quark-pion coupling due to
the off-shell effects of the quark propagator, as the
momentum of the $u-$quark  in the Born term gets  large for small 
$q^{2}$, the value  of $f^{+}(q^{2})$ would be suppressed for small
 $q^{2}$. This explains  the small value of $f^{+}(0)$ for  semileptonic
 $B$ decay. For $f_{+}(0)$ the QCD sum rule calculations give 
  $0.23\pm 0.02$\cite{Narison} close to the values $0.24\pm 0.02$
and $ 0.26\pm 0.025$ \cite{Ball2, Ball3}. The  QCDSF Collaboration lattice  
calculation\cite{QCDSF} also obtained  $f_{+}(0)=0.27\pm 0.07\pm 0.07$. 
These calculations show that the  $B\to \pi$ form factor is strongly 
suppressed at $q^{2}=0$. In term of $f^{+}(0)$ , we have: 

\bea
 &&  f_{+}(q^{2})_{D\pi}= \frac{f_{+}(0)_{D\pi}}
{(1 -q^{2}/(m_{D}^{2}+m_{\pi}^{2}))}\nonumber\\
 &&  f_{+}(q^{2})_{DK}= \frac{f_{+}(0)_{DK}}
{(1 -q^{2}/(m_{D}^{2}+m_{K}^{2}))}\nonumber\\
&&  f_{+}(q^{2})_{B\pi}= \frac{f_{+}(0)_{B\pi}}
{(1 -q^{2}/(m_{B}^{2}+m_{\pi}^{2}))}\nonumber\\
\label{ffD}
\eea
for the Born term contribution to the form factor. These are
essentially the same  to those  used in the parametrization
of the form factors  measured at BaBar, Belle and BESIII.  
 As the Born term is of pure kinematic origin, there is no
 $D^{*}, B^{*}$  pole term in our expression. This explains the fact that the 
single-pole fits for $D\to K,\pi$ form factors does not have a $D^{*}$ 
pole, consistent with the quark propagator pole term.
For the $B\to \pi$ form factor, the  mass difference between $B^{*}$
and $B$ is $45.78 \pm 0.35 \rm MeV$, the difference is negligible
for the term $q^{2}/m_{B}^{2}$ and $q^{2}/m_{B^{*}}^{2}$, one could
 then replace $m_{B^{*}}^{2}$ by $m_{B}^{2}$ witout affecting the 
BaBar BK fit,    thus making the BK parametrization consistent with 
the Born term. What is new in this paper is that the Born term could 
generate this  $q^{2}$-dependence which seems impossible to obtain otherwise. 
As shown in FIG.\ref{fig2} and FIG.\ref{fig3} for $D^{0}\to \pi^{-}$ and
$D^{+}\to \pi^{0}$ form factor, the Born term  plotted in the upper 
curve is slightly above the lower curve obtained with
the BESIII fit\cite{BES} obtained with $f_{+}^{\pi}(0)= 0.6372\pm 0.0080\pm 0.0044$,
 $M_{pole}=1.911\pm 0.012\pm 0.004{\rm GeV}$ for $D^{0}\to \pi^{-}$ and 
 $ f_{+}^{\pi}(0)= 0.6631\pm 0.092\pm 0.0041$,
 $M_{pole}=1.898\pm 0.020\pm 0.003{\rm GeV}$
for   $D^{+}\to \pi^{0}$ form factors respectively,  while the middle curve, obtained by adding a small polynomial term to the Born term, now
almost coincides with the lower curve.A similar small polynomial term
is also needed to fit the $D^{+}\to \pi^{0}$ form factor BESIII data
shown in FIG.\ref{fig3}\cite{BES2}. The Belle data\cite{Belle}, as mentioned 
earlier, obtained a single pole mass $m_{\rm pole}(K^{-}\ell^{+}\nu)=
1.82\pm 0.04\pm 0.03 \rm GeV$ for $D^{-}\to K^{-}\ell^{+}\nu$ decay,
lower than the value of $1.935\rm GeV$ of BESIII data, while for 
$D^{+}\to \pi^{0}$ decay, $m_{\rm pole}(\pi^{-}\ell^{+}\nu)=
1.97\pm 0.08_{\rm stat}\pm 0.04_{\rm syst} \rm GeV$ is slightly above
 the BESIII value of $1.898\pm 0.020\pm 0.003 \rm GeV $
BK fit with $\alpha= 0.52\pm 0.08_{\rm stat}\pm 0.06_{\rm syst}$, while
for the $D^{+}\to \pi^{0}$, a similar BK fit with $\alpha= 0.10\pm
0.21_{\rm stat}\pm 0.10_{\rm syst}$, making the fit close to the single
 pole model. This is consistent with a small correction we have for 
the $D^{+}\to \pi^{0}$ form factor shown in FIG. 2. In the BaBar 
measurements, the simple pole mass for the $D^{0}\to \pi^{-}$ form factor
is $1.906 \pm 0.029\pm 0.023) $ very close to the BESIII value of 
$1.911 \pm 0.012\pm 0.004) $, for $D^{0}\to K^{-}$, the pole mass is 
$1.884 \pm 0.012\pm 0.016) $ rather below the BESIII
value of $1.921 \pm 0.010\pm 0.007) $. The  Fermilab Lattice Collaboration,
MILC Collaboration and HPQCD Collaboration,  three-flavor QCD
Calculations\cite{MILC} also have lattice results for $D\to \pi$ and $D\to K$
fitted with BK parametrization with $B^{*}$ pole and obtain results for  
the CKM matrix elements in agreement with experiments.
\begin{figure}[ht]
\centering
\includegraphics[height=5cm,angle=0]{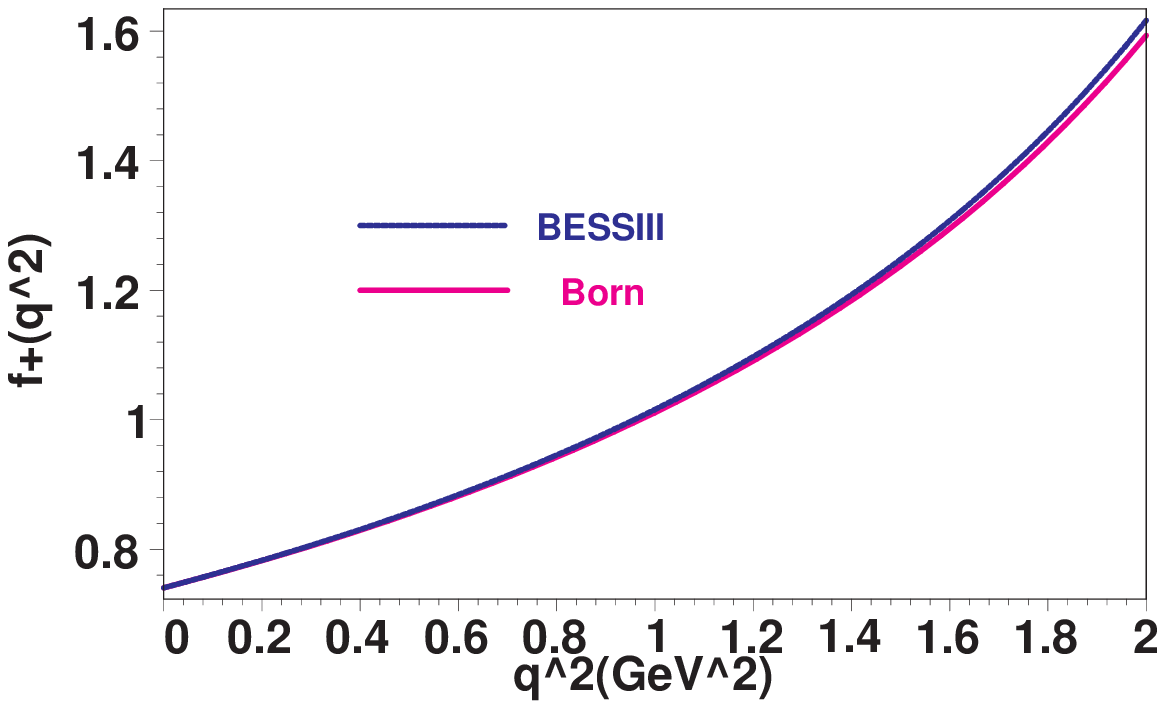}
\caption{The Born term(lower curve) and the fit to the  $D^{0}\to K^{-}$ BES 
measured form factor (upper curve)  given in FIG.8 of Ref. \cite{BES} }
\label{fig4}
\end{figure}
\begin{figure}[ht]
\centering
\includegraphics[height=5cm,angle=0]{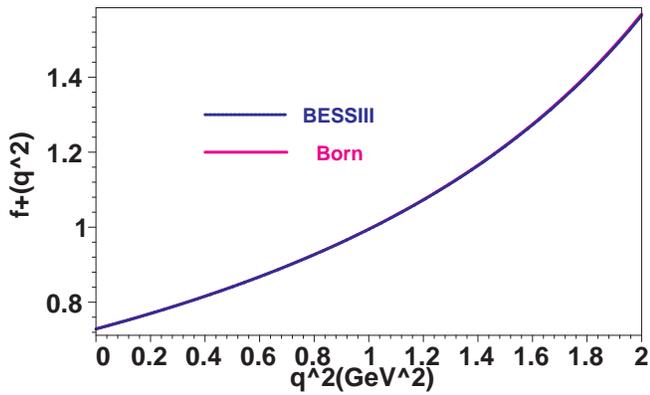}
\caption{The Born term(lower curve) and the fit to the  $D^{+}\to \bar{K^{0}}$ BES  measured form factor  (upper curve)  given in FIG.6 of Ref. \cite{BES2}}
\label{fig5}
\end{figure}

 For the $D^{0}\to K^{-}$ form factor, as shown in FIG. \ref{fig4}, the lower
curve(Born term)  is in excelllent agreement with
the fit to the BESIII  data(upper curve) \cite{BES} with 
$f_{+}^{K}(0)= 0.7768\pm 0.0026\pm 0.0036$,
$M_{pole}=1.921\pm 0.010\pm 0.007\rm GeV$ $V_{cs}=0.97343\pm 0.00015$ 
and  $V_{cs}=0.97343\pm 0.00015$. This good
agreement could be explained by the  $m_{K}^{2}$ term in the $u-$quark 
propagator. If we replace the  factor $q^{2}/(m_{D}^{2} + m_{K}^{2}) $ by 
$q^{2}/m_{\rm eff}^{2}$, with $m_{\rm eff}= \sqrt{m_{D}^{2} + m_{K}^{2}} $
as the effective mass in the pole term, then
$m_{\rm eff}=1.931{\rm GeV} $, very close to the pole mass of the BESIII 
fit\cite{BES},  $M_{pole}=1.921\pm 0.010 \pm 0.007{\rm GeV}$. This 
explains the agreement between  the two curves in FIG. \ref{fig4}.
Agreement  is also found between the Born term and the BESIII 
fit\cite{BES2}  for the $D^{+}\to \bar{K^{0}}$ form factor obtained with 
$f_{+}^{\pi}(0)=0.7094\pm 0.0035\pm  0.0111$, 
$M_{pole}=1.935\pm 0.017\pm 0.006{\rm  GeV}$ shown in FIG.\ref{fig5} , 
very close to the  effective mass $m_{\rm eff}=1.931{\rm GeV} $ in the
 Born term. This dependence on $m_{K}^{2} $ in both $D^{0}\to K^{-}$ and 
$D^{+}\to\bar{K^{0}}$ form factor shows evidence for the dominance
of the Born term for the $D\to K$ semileptonic decay form factors.

\begin{figure}[ht]
\centering
\includegraphics[height=5cm,angle=0]{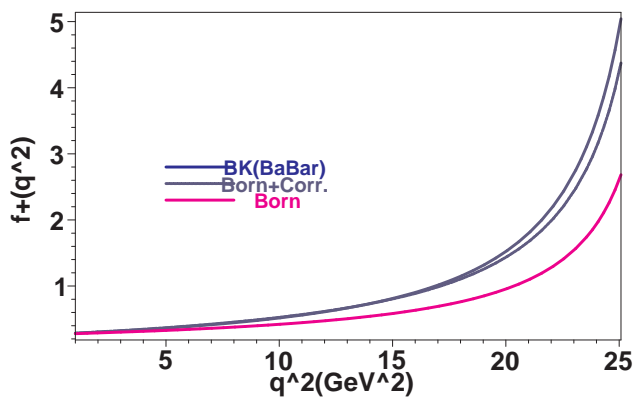}
\caption{The Born term(lower curve), the Born+Corr plot has  a small
polynomial term added (middle curve) and  the BaBar data (upper curve)
represented by  the BK fit to the BaBar data of Ref. \cite{BaBar1}}
\label{fig6}
\end{figure}
  For the $B\to \pi$  form factor, there is also a correction to the 
Born term  to compensate for a suppression at large $q^{2}$ induced by
$f_{+}(0)_{B\pi}$ mentioned above. The reason is that, for small $q^{2}$,
the quark in the Born term quark propagator is highly virtual and is far
off the mass shell, the pion-quark coupling has a suppressed form factor,
 making the $B\to \pi$ or $D\to \pi$ form factor suppressed by the suppression
of $f_{+}(0)_{B\pi}$. To compensate for this suppression, a correction term
 $(1 + a q^{2}/(m_{B}^{2}+m_{\pi}^{2}))$ for $B\to \pi$
is needed to bring  the form factor in agreement with data at large
$q^{2}$. 

 With these corrections included, the  middle curve of
FIG. \ref{fig6} is  now in agreement
with data and    almost  coincides with the lower curve obtained with
a BK fitted to the BaBar data\cite{BaBar1}. Thus the  $D\to K,\pi$ and 
$B\to \pi$ form  factors  with the Born term as  the main  contribution, 
and assuming  the same correction term for $D^{0}\to \pi^{-}$ and
$D^{+}\to \pi^{0}$ form factor, are now  in agreement with  data and are
given by~:

\bea
 &&  f_{+}(q^{2})_{D\pi}= \frac{f_{+}(0)_{D\pi}(1 -0.15 q^{2}/(m_{D}^{2}+m_{\pi}^{2}))}{(1 -q^{2}/(m_{D}^{2}+m_{\pi}^{2}))}\nonumber\\
 &&  f_{+}(q^{2})_{DK}= \frac{f_{+}(0)_{DK}}
{(1 -q^{2}/(m_{D}^{2}+m_{K}^{2}))}\nonumber\\
 &&  f_{+}(q^{2})_{B\pi}= \frac{f_{+}(0)_{B\pi}(1 + 0.70 q^{2}/(m_{B}^{2}+m_{\pi}^{2}))}{(1 -q^{2}/(m_{B}^{2}+m_{\pi}^{2}))}\nonumber\\
\label{ffDc}
\eea

\section{Conclusion}
  
  In conclusion, we have shown that, the tree-level Born term for the
 process $c+ {\bar d}\to \pi  \ell \nu$ and $b+ {\bar d}\to \pi \ell \nu$
in semileptonic decays of a heavy meson to a light meson in the final
state  is found to describe rather well the $q^{2}$-dependence of the
$D\to \pi$,$D\to K$ and $B\to \pi$ form factors. In particular, the
$D^{0}\to K^{-}$ and $D^{+}\to\bar{K^{0}}$ form factors show possile
evidence for the $K$ meson mass term  in the quark propagator and  the
simple  $q^{2}$-dependence generated by this Born term.

\bigskip
\noindent $\bullet$ Note added: After completion of this paper, I found a 
previous paper\cite{Riazuddin} in which a B-meson pole
$q^{2}$-dependence term $1/(1-q^{2}/m_{B}^{2})$  for the $B\to \pi$ form 
factor is also obtained.

\end{document}